\documentclass[a4paper,11pt]{article}
\usepackage{jinstpub} 

\usepackage[english]{babel} 
\usepackage{placeins}  % For floatbarrier
\usepackage[per-mode=symbol, uncertainty-mode = separate]{siunitx}
\DeclareSIUnit\bar{bar}
\DeclareSIUnit\electron{e^-}

\addto\extrasenglish{}
\addto\extrasenglish{}

\title{Charge carrier generation in RNDR-DEPFET detectors}

\author[a,b,c ~]{Niels Wernicke}
\author[d ~]{Alexander Bähr}
\author[a,c ~]{Hannah Danhel}
\author[a,c ~]{Florian Heinrich}
\author[a ~]{Holger Kluck}
\author[d ~]{Jelena Ninkovic}
\author[a,c ~]{Jochen Schieck}
\author[a,b ~]{Wolfgang Treberspurg}
\author[d ~]{Johannes Treis}

\affiliation[a]{Marietta Blau Institute for Particle Physics (MBI), Dominikanerbastei 16,  3rd floor, 1010 Vienna, Austria}
\affiliation[b]{Fachhochschule Wiener Neustadt, Johannes Gutenberg-Straße 3, 2700 Wiener Neustadt, Austria}
\affiliation[c]{TU Wien, Wiedner Hauptstr. 8–10, 1040 Vienna, Austria}
\affiliation[d]{Halbleiterlabor der Max-Planck-Gesellschaft (HLL), Isarauenweg 1, 85748 Garching, Germany}
\emailAdd{niels.wernicke@oeaw.ac.at}
\emailAdd{wolfgang.treberspurg@oeaw.ac.at}

\abstract{
Depleted p-channel field-effect transistor detectors with repetitive non-destructive readout (RNDR-DEPFETs) achieve deep sub-electron noise by averaging several independent measurements of a single event. During repetitive readout, collected electrons are transferred between two readout nodes within each pixel to enable electron number-resolved measurements. The pixels serve as the unit cells of an active pixel sensor, enabling a high level of parallelization and fast readout. These properties are exploited in the DANAE experiment, which aims at the direct detection of light dark matter using the signature of electron recoils.

We present the experimental characterization of a $64\times64$ RNDR-DEPFET pixel detector, with a focus on the charge-carrier generation rate. This technology provides high time resolution, which improves sensitivity to rare events involving signals of two or more electrons. This follows from the Poisson statistics of thermally generated electrons. 
}

\keywords{ % https://jinst.sissa.it/jinst/help/helpLoader.jsp?pgType=kwList
Dark Matter detectors (WIMPs, axions, etc.), Particle detectors, Ionization and excitation processes, Thermal noise
} 
\arxivnumber{7000954}

\begin{document}
\maketitle
\flushbottom

\section{Introduction}
Dark matter (DM) is an integral constituent to explain the movement of stars and galaxies in the universe~\cite{DMRotation} but it has only been observed indirectly. However, direct detection experiments excluded a large parameter space especially for heavy DM candidates. This triggers the need for experiments that are sensitive in not excluded regions~\cite{cosinusAngloher} as well as an increasing interest on light DM. 

To be sensitive to light DM, the event signature of the scattering of DM candidates on electrons is studied instead of the scattering process on target nuclei~\cite{EssigDirect}. The target and, at the same time, sensing material of the DANAE experiment is silicon. Single electrons, that interact with DM candidates are excited into the conduction band of the semiconductor by a minimum energy transfer of \SI{1,12}{\electronvolt} --- the bandgap of silicon. The excited electrons are collected, stored and read out to measure the deposited energy. To minimize background signals from thermal excitation and environmental radiation, the device has to be cooled down to temperatures of about \SI{140}{\kelvin} and needs to be shielded from radiation sources.

\section{The RNDR-DEPFET principle}
\begin{figure}[htb] 
    \centering 
    \includegraphics[width=0.6\linewidth]{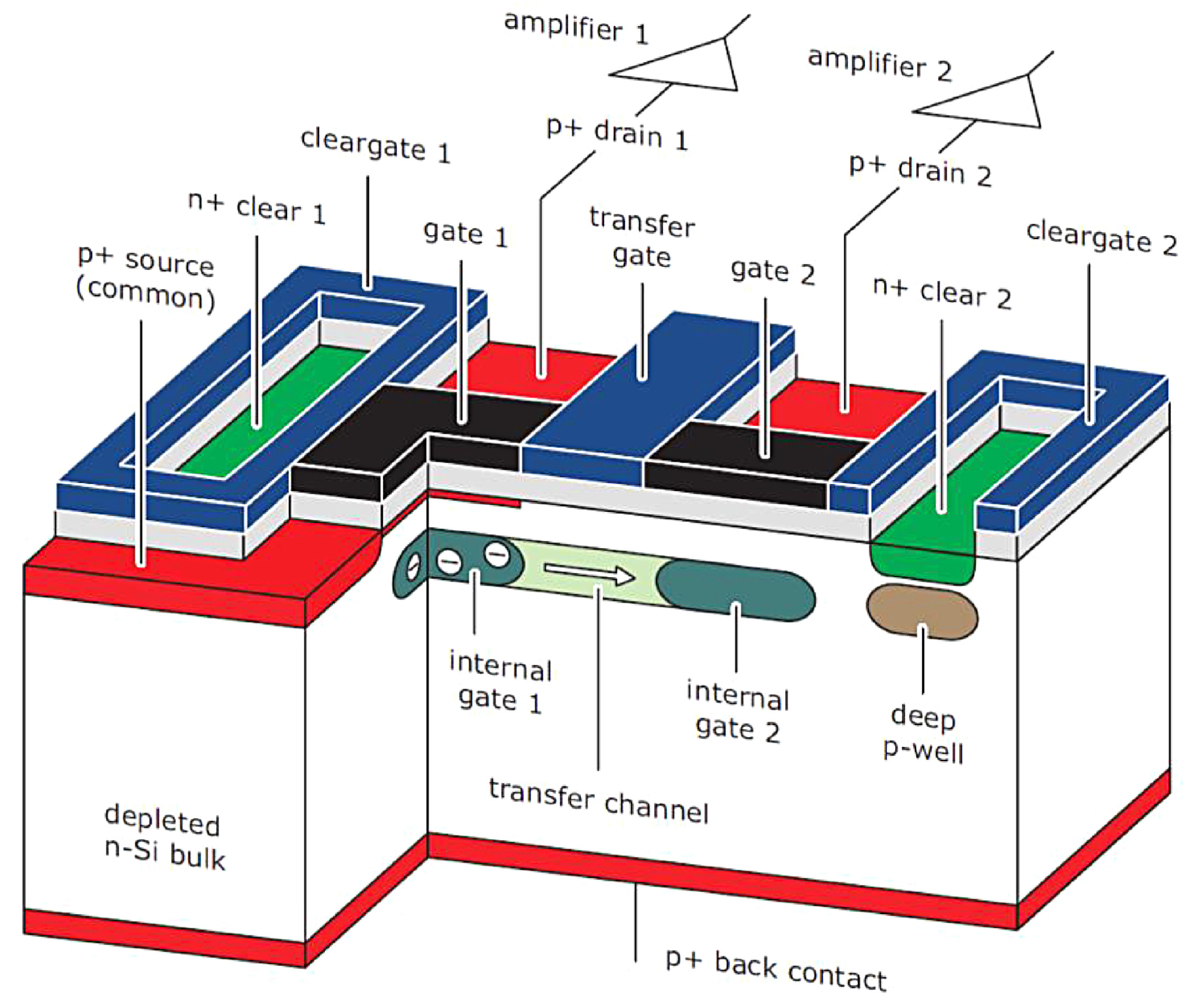} 
    \caption{RNDR-DEPFET schematic: Electrons are collected in the internal gate to modify the conductivity of the transistor channel. An RNDR-DEPFET pixel hosts two sub-pixels to interchange electrons for repetitive readout~\cite{DEPFETBaehr}.}
    \label{fig:RNDR_DEPFET} 
\end{figure}

The DEPFET detector includes a fully depleted semiconductor sensor (\autoref{fig:RNDR_DEPFET}). As an active pixel sensor (APS), the first stage of amplification is implemented in each pixel~\cite{Lauf}. Charge carriers that are generated in the depleted bulk drift to the minimum of the electrical potential, which is realized with a deep n-implant located below the transistor gate. Inside this so-called internal gate, the collected electrons induce mirror charges in the transistor channel and thus modify its conductivity. This way a measurement of the amount of collected electrons is realized without disturbing them~\cite{Lauf}. After measuring the channel conductivity with the internal gate filled, the electrons are removed from the internal gate using either a clear gate, a clear structure, or a transfer gate. The DEPFET baseline is then measured with an empty internal gate and subtracted from the previous measurement to implement \textit{correlated double sampling} (CDS). With a single CDS measurement, the signal can be determined with an uncertainty of about 5\,$e^-$~\cite{DEPFETBaehr}, which is not precise enough for electron number resolution. 

For this reason, a RNDR-DEPFET pixel combines two DEPFET structures connected via a transfer gate (\autoref{fig:RNDR_DEPFET}). The charge carriers are shifted from the internal gate of one DEPFET sub-pixel to the other to enable multiple independent measurements before they are finally cleared~\cite{DEPFETBaehr}. Charge carriers are transferred by applying a voltage on the transfer gate, which connects the internal gates of both sub-pixels. The electrical potential of the internal gate of each sub-pixel is coupled to the corresponding external gate. Thus, as soon as the transfer-gate opens, electrons drift from the sub-pixel, that is turned on to the sub-pixel that is turned off by the external gate. This way, each sub-pixel is read out with a CDS process. Afterwards, the electrons are moved in the opposite direction to facilitate multiple independent measurements of the generated charge carriers. 

\section{The DANAE detector}
The prototype detector of the DANAE (direct dark matter search using DEPFET with repetitive non-destructive readout application experiment) project operates a $64\times64$ pixel sensor. Each pixel includes an RNDR-DEPFET with a thickness of \SI{450}{\micro\meter} and a side length of \SI{50}{\micro\meter}, leading to a target mass of the sensor of \SI{10,7}{\milli\gram}~\cite{baehrSize}. The sensor is operated in a rolling-shutter mode --- the 64 pixels of one row are read out consecutively in parallel --- and the frame rate of a $64\times64$ pixel detector is approximately one second with sub-electron resolution~~\cite{DEPFETBaehr}. The subsequent measurements were taken with 800 repetitions, which results in \SI{1,22}{\second} per frame. The multiplexing of the 64 in parallel signals sampled is done after the CDS in a simplex mode. Thus, the achievable time resolution can be increased by a factor of two in a duplex operation.

The operating temperature is approximately \SI{140}{\kelvin} in a vacuum of about \SI{5e-6}{\milli\bar}. The heat shielding as well as the vacuum chamber are made out of aluminium. With the current location and shielding at the ground floor in a conventional laboratory ($\sim\SI{190}{\meter}$ above sea level,  $\sim \SI{1,2}{\meter}$ of concrete from the floors above,  $\sim\SI{1}{\centi\meter}$ of stainless steel from the vacuum chamber, and $\sim\SI{5}{\centi\meter}$ of aluminium from the heat shielding) the amount of background radiation can be compared to surface runs of other experiments. In the long term, the experiment is intended to move to a deep underground laboratory, such as the \textit{Laboratori Nazionali del Gran Sasso} (LNGS).

\section{Spectral resolution of DANAE}
To discriminate between single electrons and to be able to study the single-electron event (SEE) rate, a proper spectral resolution is necessary. The signal resolution is dominated by the readout noise, which in turn decreases with the number of averaged readings. In the absence of any disturbances such as shot noise or transfer inefficiencies, $N$ measurements reduce the noise by $1/\sqrt{N}$. For the data acquisition, a readout setting with 800 repetitions turned out to be best suited to discriminate between the electron peak while minimizing signal loss due to transfer inefficiencies. 

The pixels of each row are read out in parallel, the readout time for one pixel is the same as the readout time for the whole row. To read out a pixel, the charges are first transferred to the internal gate of sub-pixel~1. After a short settling time, the first measurement signal integration takes place, followed by the transfer of charge carriers to the internal gate of sub-pixel~2 before the second signal integration for a CDS is performed.  For each subpixel, the CDS difference between the charge measurement and the empty-gate baseline is formed and multiplexed to the ADC. The same process is repeated in reverse order to start the second signal sampling.

\begin{figure}[htb]
    \centering
    \includegraphics[width=0.70\linewidth]{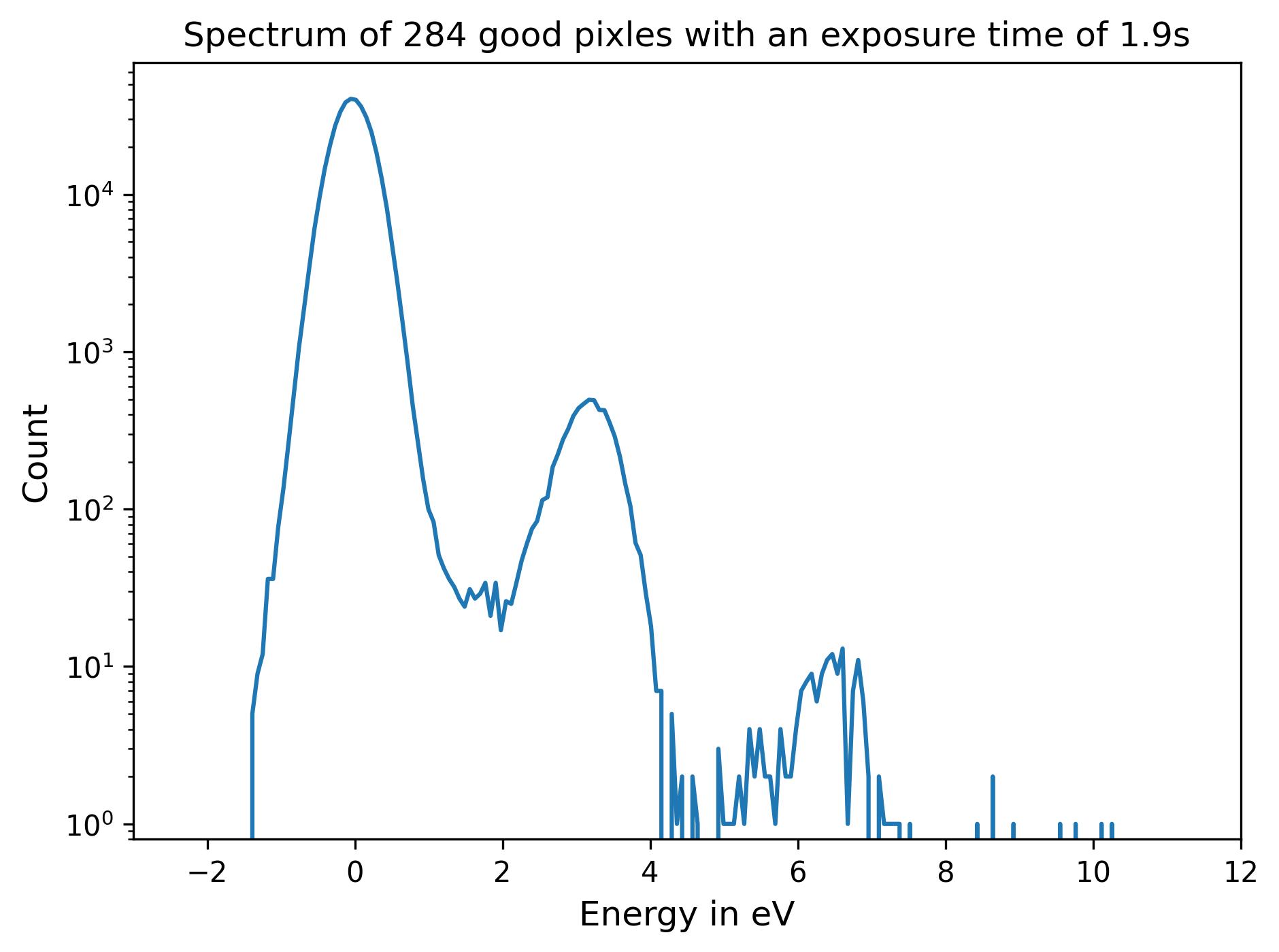}
    \caption{Recorded raw data spectrum for 110 pixels after 800 repetitions and a single sampling of \SI{1,9}{\second}.}
    \label{fig:NREP}
\end{figure}

For each repetition, the multiplexing to the ADC takes \SI{6,4}{\micro \second}, two integrations of \SI{8}{\micro \second} each, and one transfer and one settling time of \SI{400}{\nano \second} each are required. In total, the readout of one repetition of one row of pixels takes about \SI{23}{\micro\second}, and  800 repetitions sum up to \SI{38}{\milli\second}. Afterwards, free electrons are cleared from the row in \SI{6,4}{\micro \second}. Then the next row is read out. Without additional exposure time, the readout of the 64 rows of the sensor takes about \SI{1,2}{\second}.
The raw-data spectrum is a histogram of all measured offset and common-mode-corrected values for all pixels and all frames. The raw-data spectrum for a set of 184\,pixels demonstrates the capability to distinguish single-electron signals (\autoref{fig:NREP}).

\section{Data acquisition and filtering}\label{sec:DataFilter}
The readout time of one frame of \SI{1,22}{\second} for 800 repetitions is the minimal achievable exposure time. To probe the generation rate, an additional exposure between \SI{100}{\milli\second} and \SI{4,0}{\second} was added after the last line of each frame. All other variables, such as the operating voltages and the timing of the operation sequence, remain unchanged for different exposures. The temperature drift of the sensor is less than \SI{0,5}{\kelvin}. 

\begin{figure}[htb]
    \centering
    \includegraphics[height = 60mm]{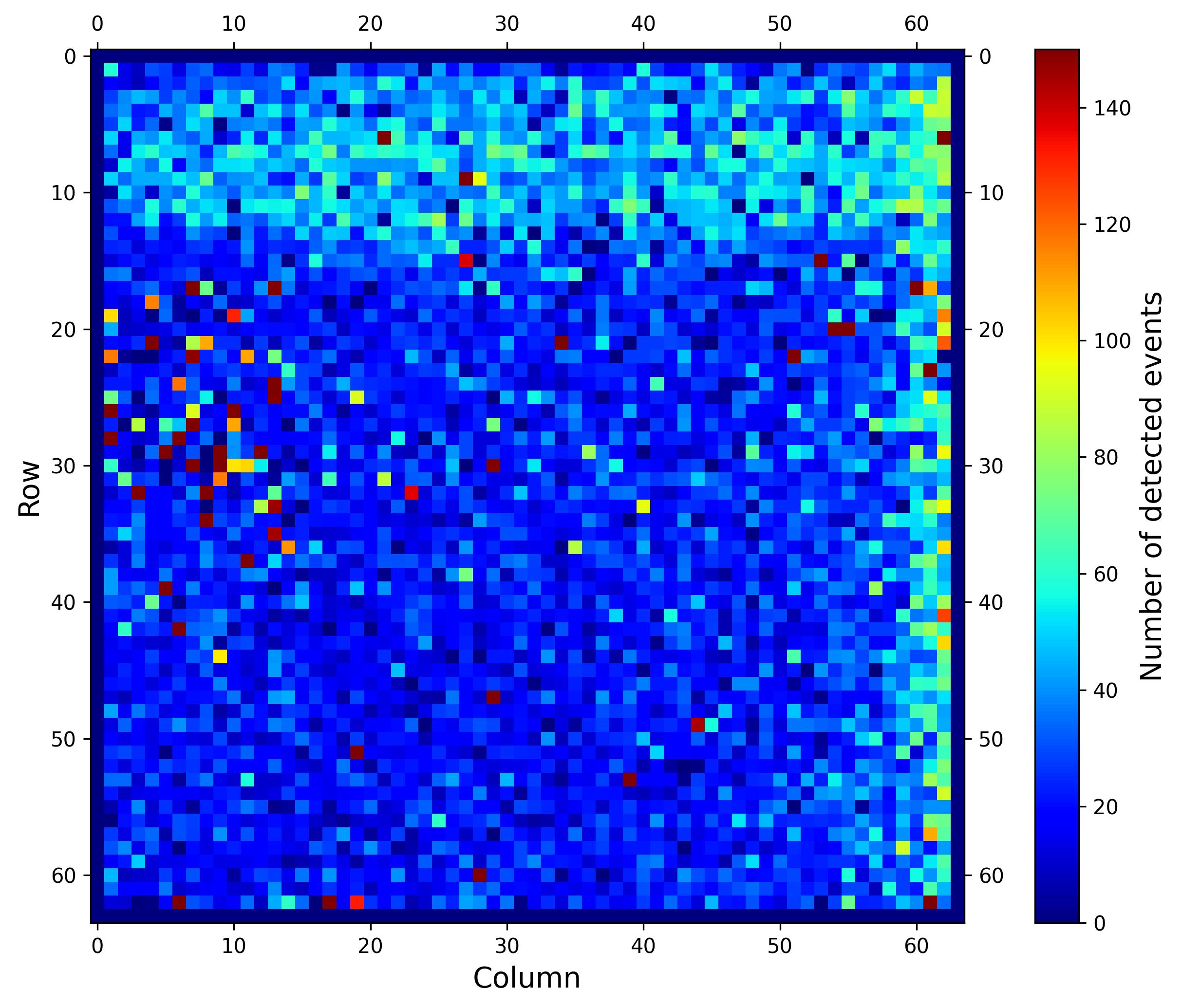}
    \caption{The hit map for a total exposure time of \SI{1,9}{\second}.}
    \label{fig:EventMap}
\end{figure}

For event filtering, a threshold of five sigma was applied to the spectra of each non-calibrated pixel (\autoref{fig:SinglePixle}). A hit map shows, for each pixel, the number of frames in which its value exceeds the Gaussian expectation by more than five sigma. For an exposure run with \SI{1,9}{\second}, such a map reveals a homogeneous distribution with the exception of the sensor edges (\autoref{fig:EventMap}).

About 1400 frames are taken for each measured exposure run. Incomplete frames due to the analogue-to-digital converter (ADC) settings are dropped prior to the analysis. The data are analysed with the custom-developed, Python-based APANTIAS software (version 2.0.3). To identify a representative set of pixels, several constraints were applied to filter out affected signals and pixels:

\paragraph{Border pixels}
The pixels at the border of the sensor are affected by the electrical-field configuration at the edges and are discarded. Because they cannot be adequately shielded from the surrounding substrate, charge sharing affects them differently. 

\begin{figure}[htb]
    \centering
    \includegraphics[width=0.9\linewidth]{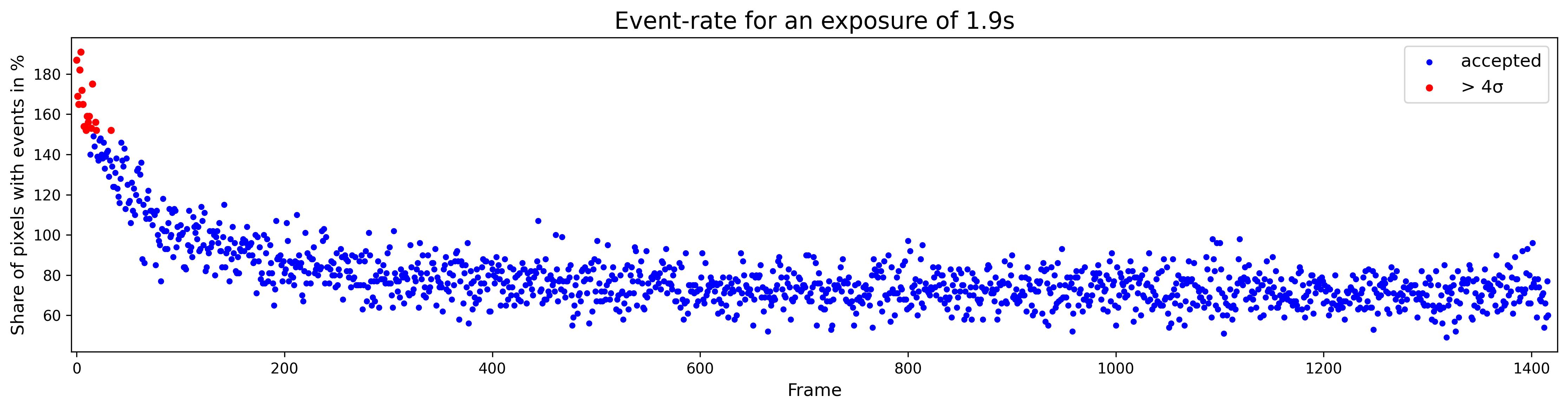}
    \caption{The percentage of pixels with events are plotted against the number of frames. The red values are removed.}
    \label{fig:EventsPerFrame}
\end{figure}

\paragraph{Bad frames}
Frames for which the average event number exceeds the median of all recorded frames by more than five sigma are removed (\autoref{fig:EventsPerFrame}). Most of the removed frames occur at the beginning of the measurement. This is because of the configuration of a new measurement sequence, which takes about a minute. During this time, charge carriers are generated. The charge carriers generated during the programming of the detector are not cleared after the first frame. Thus, frames recorded until the sensor settles need to be dismissed.

\begin{figure}[htb] 
    \centering
    \includegraphics[height = 50mm]{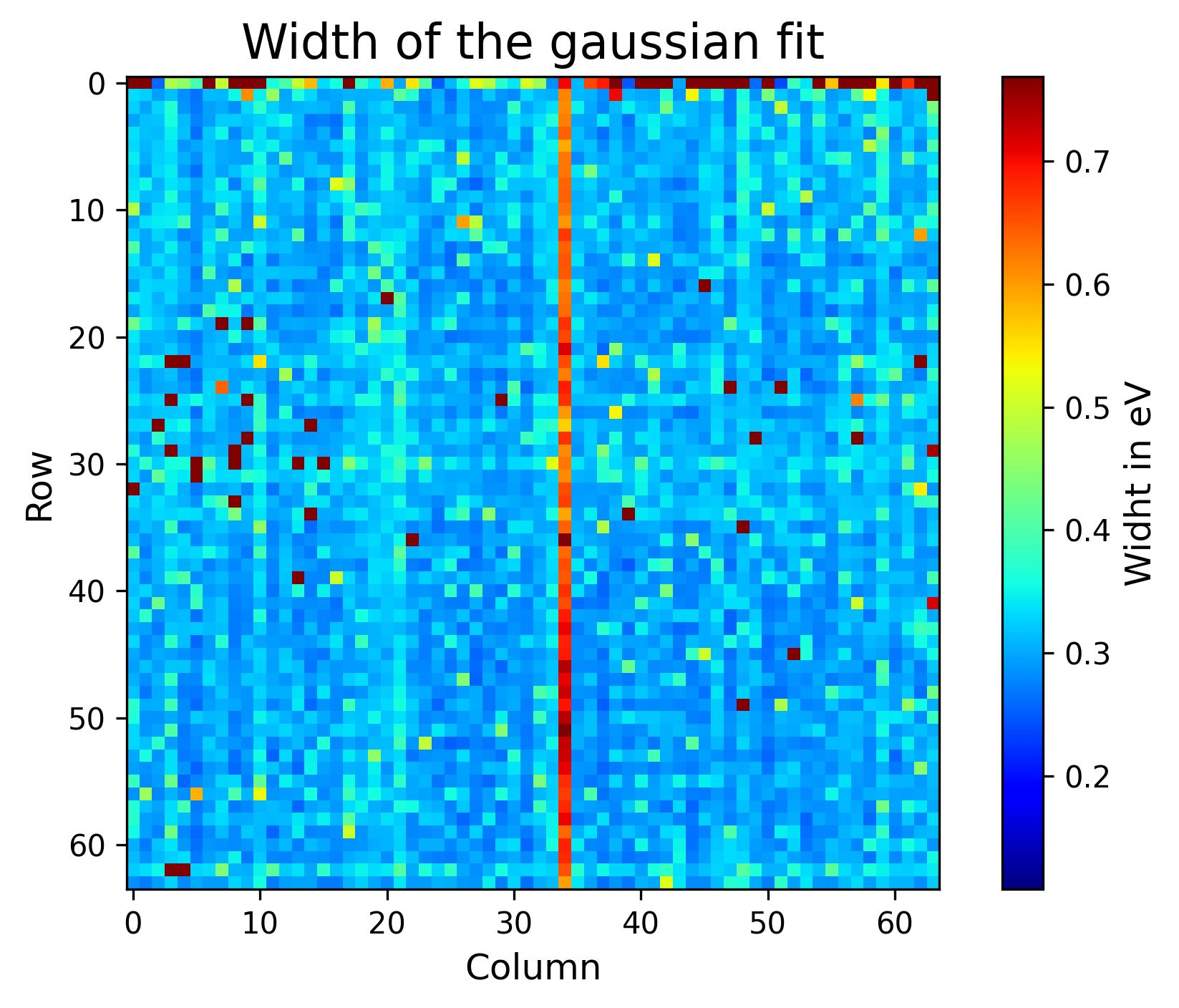} 
    \qquad
    \includegraphics[height = 50mm]{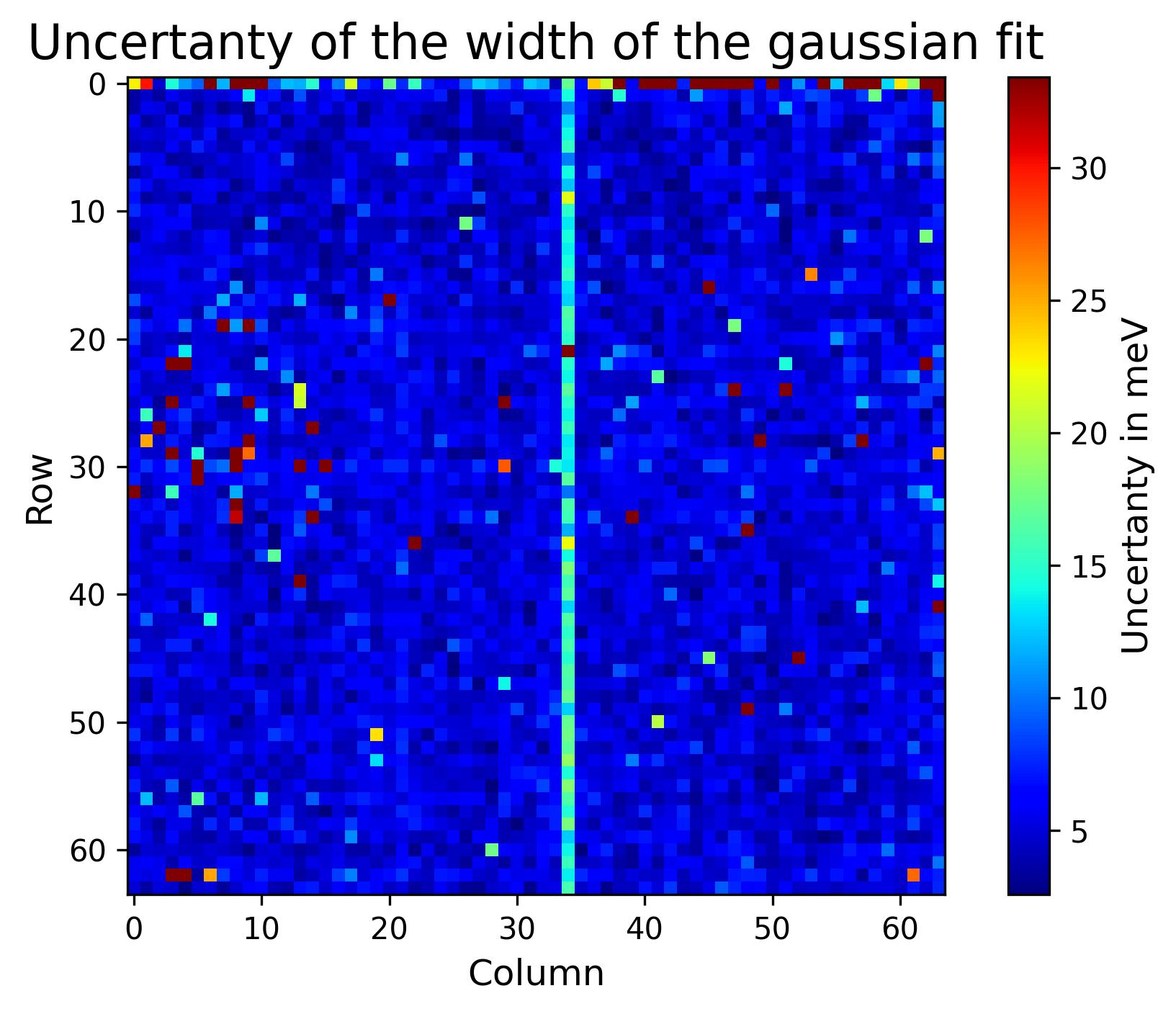} 
    \caption{The width of the Gaussian fit and its uncertainty for an exposure time of \SI{1,9}{\second}. For column~34, the width of the Gaussian fit is abnormally large, indicating excess noise. This is likely due to defects in this readout channel. } 
    \label{fig:NoisePixels}
\end{figure}

\paragraph{Bad rows and columns}
The noise determination is done with a a Gaussian fit to the histogram of averaged readings per pixel. Due to the irregularities in column~34, this column was discarded (\autoref{fig:NoisePixels}).  The standard deviation of the Gaussian is on average \SI{0,31}{\electronvolt} across the sensor, while for this column it is \SI{1,08}{\electronvolt}. This is the case at all exposure times, indicating a hardware problem in this readout channel. As a consequence, these pixels are rejected from further analysis.

\begin{figure}[htb]
    \centering
    \includegraphics[height = 50mm]{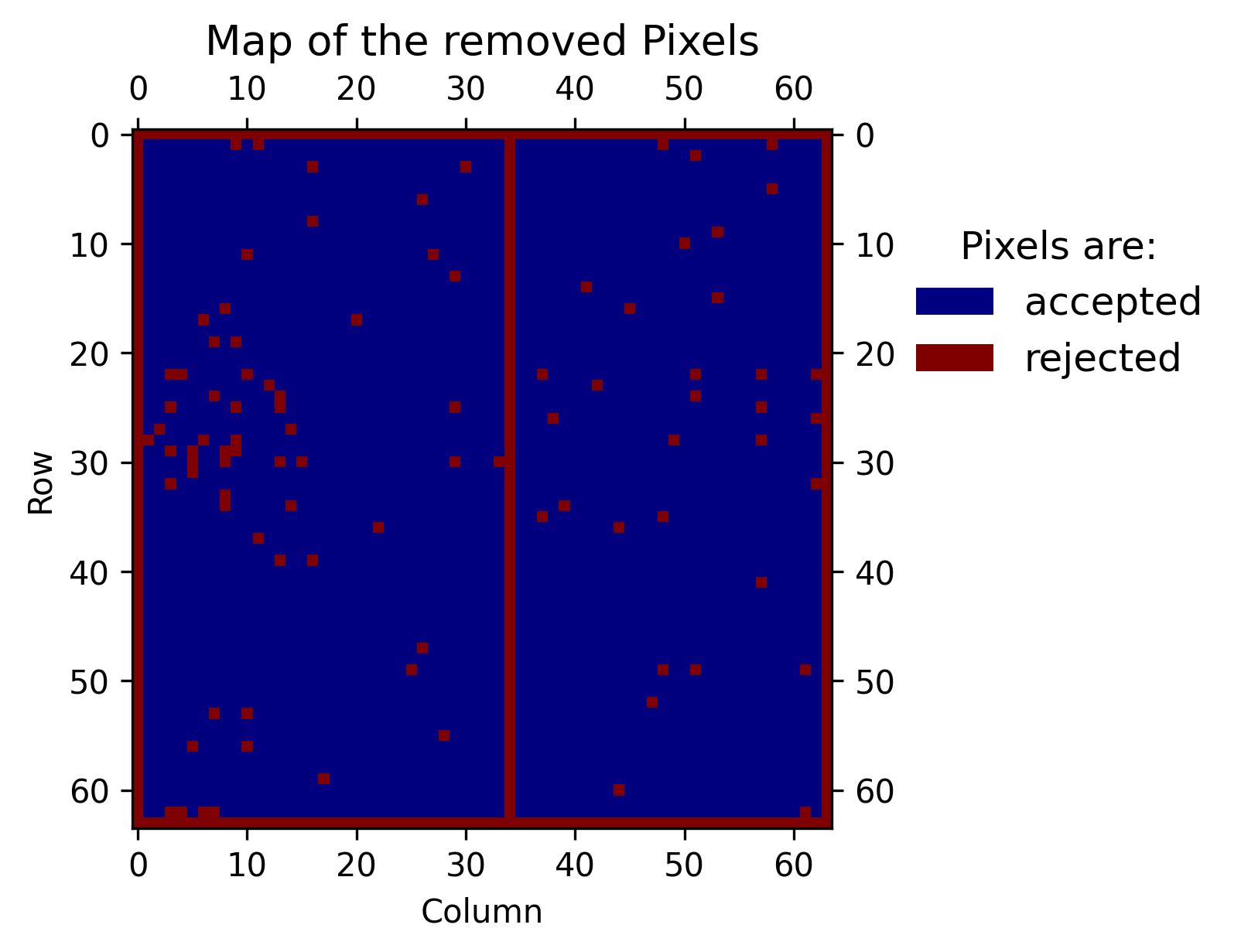}
    \caption{The mask pixels after all the cuts in \autoref{sec:DataFilter} and \autoref{sec:DataSpektrum}. Of the 3844 pixels that are not at the edge of the sensor, 154 are blinded (4.0\%).}
    \label{fig:MaskPixle}
\end{figure}

\paragraph{Hot and cold pixels}
In bright or hot pixels, charge carriers are generated more frequently, mainly due to well localized impurities or defects. Cold pixels are less sensitive or do not detect signals at all. To identify the hot and cold pixels, the number of detected clustered hits over all frames is calculated for each pixel, resulting in 4096 values. The distribution of these values is calculated, pixels with more than five sigma difference from the median pixel value are masked. his removes an additional 20 pixels (\autoref{fig:MaskPixle}).

\paragraph{Pixel clustering}
If more than one electron is generated in adjacent pixels, the signal can spread over those pixels. In the hit map, this would appear as multiple hits that need to be assigned to one event. In the spectrum two entries at low energies are merged to one with a higher signal. This effect is handled by event clustering. This hardly impacts the generation rate, since the rate of multiple-pixel events is dominated by pile-up of thermal single-electron events and is therefore time dependent as well. 

The generation rate is not significantly impacted by treating all hits as single-electron events, since the fraction of two-electron events is less than 2\,\%, and the fraction of three- and four- electron events is less than 0.1\,\%.

\section{Raw data spectrum after filters}\label{sec:DataSpektrum}
\begin{figure}[htb]
    \centering
    \includegraphics[width = 0.95\linewidth]{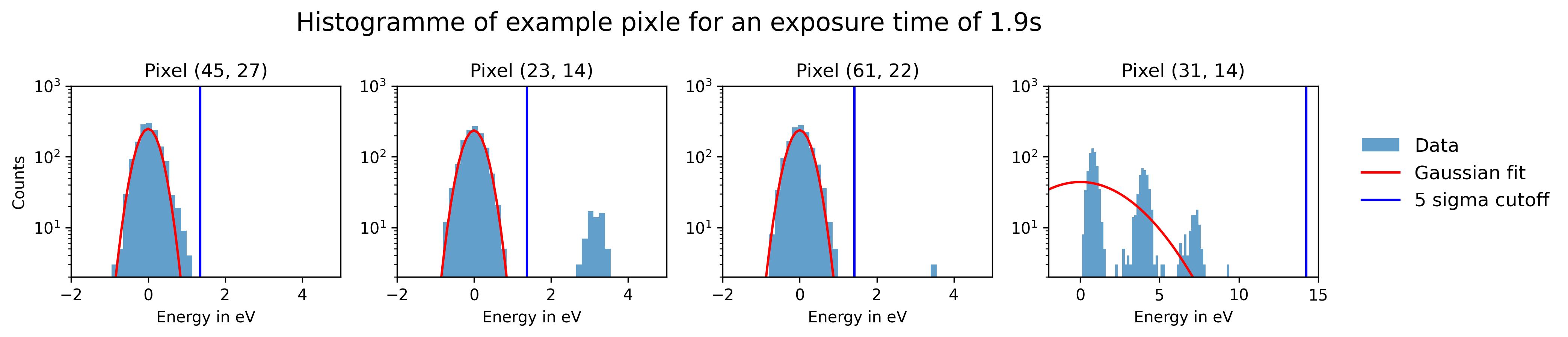}
    \caption{Spectra of the representative pixels. From left to right: in the first pixel, no events were detected. In the next two pixels, the one-electron peak is well pronounced, with a different number of events. For the last pixel, there were problems with the common-mode correction. This lead to the spectrum shown with multiple large peaks. Due to these problems, this pixel is masked and not used for the analysis. The spectra are taken with a total exposure time of \SI{1,9}{\second} per frame. A Gaussian fit is applied, and the 5-sigma threshold is shown as a blue line.}
    \label{fig:SinglePixle}
\end{figure}

The raw-data spectrum corresponds to the recorded averaged measurements accumulated over all pixels and all frames. The spectra of four representative individual pixels illustrate different pixel properties and the capability of event identification, including one pixel that is rejected from the analysis (\autoref{fig:SinglePixle}).

\begin{figure}[htb]
    \centering
    \includegraphics[height = 50mm]{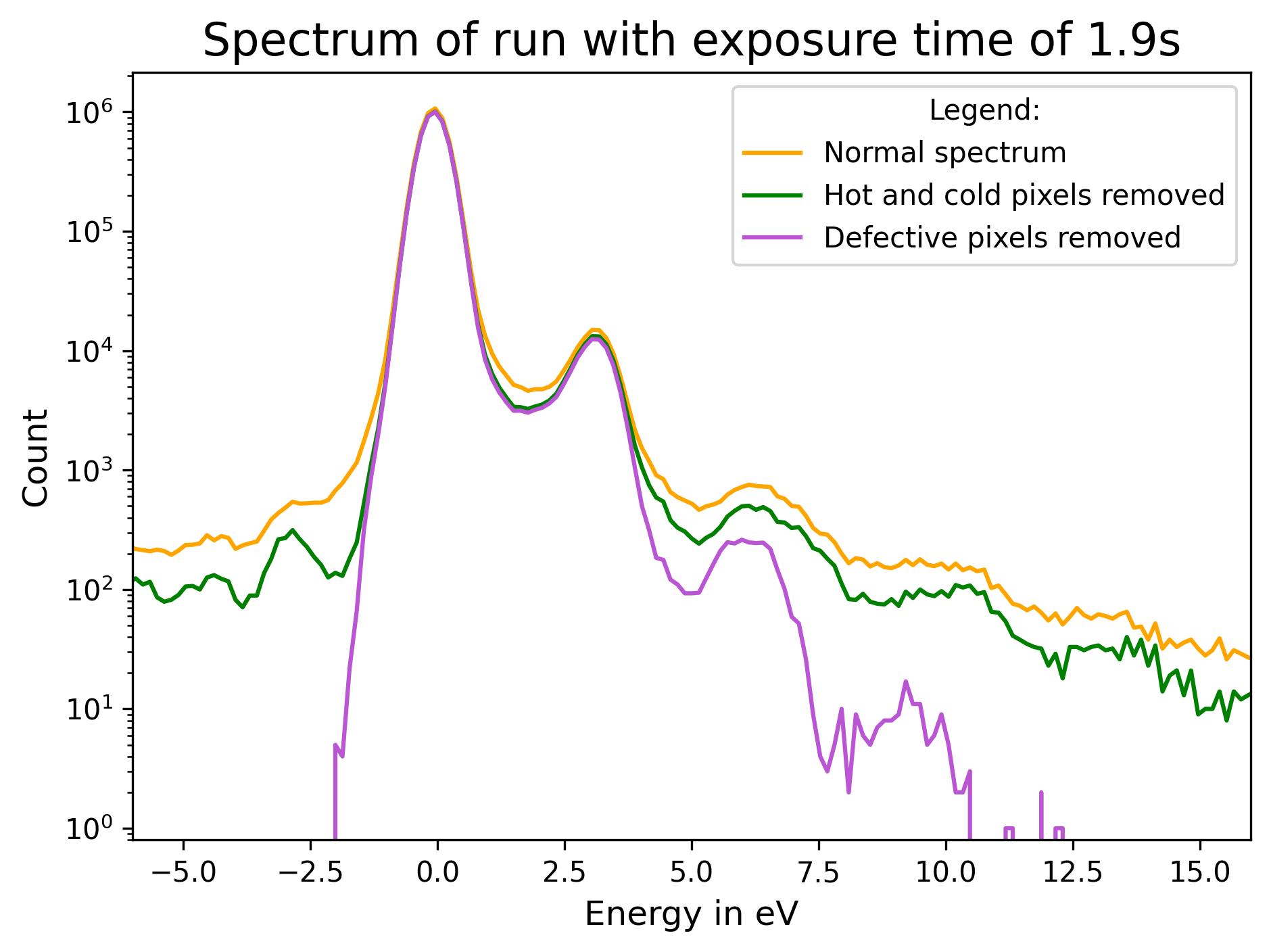}
    \caption{The hot and cold pixel and the bad rows and columns are removed for the green histogram. This leads to an improved peak-to-valley ration. Removing an additional 40 defective pixels yields the purple curve.}
    \label{fig:SpectrumMasked}
\end{figure}

Filtering defective pixels removes artificial signals in the logarithmically plotted raw-data spectrum and improves the peak-to-valley ratio (\autoref{fig:SpectrumMasked}).

Masking defective pixels removes an additional 55 pixels, leading a total cut of 406 pixels, 252 of which are at the border (\autoref{fig:MaskPixle}). In total, about 9.9\% of pixels are discarded, which is a rather permissive approach and leads to an effective active mass of \SI{9,6}{\milli\gram}

\section{Charge carrier generation rate}
\begin{figure}[htb]
    \centering
    \includegraphics[width=0.65\linewidth]{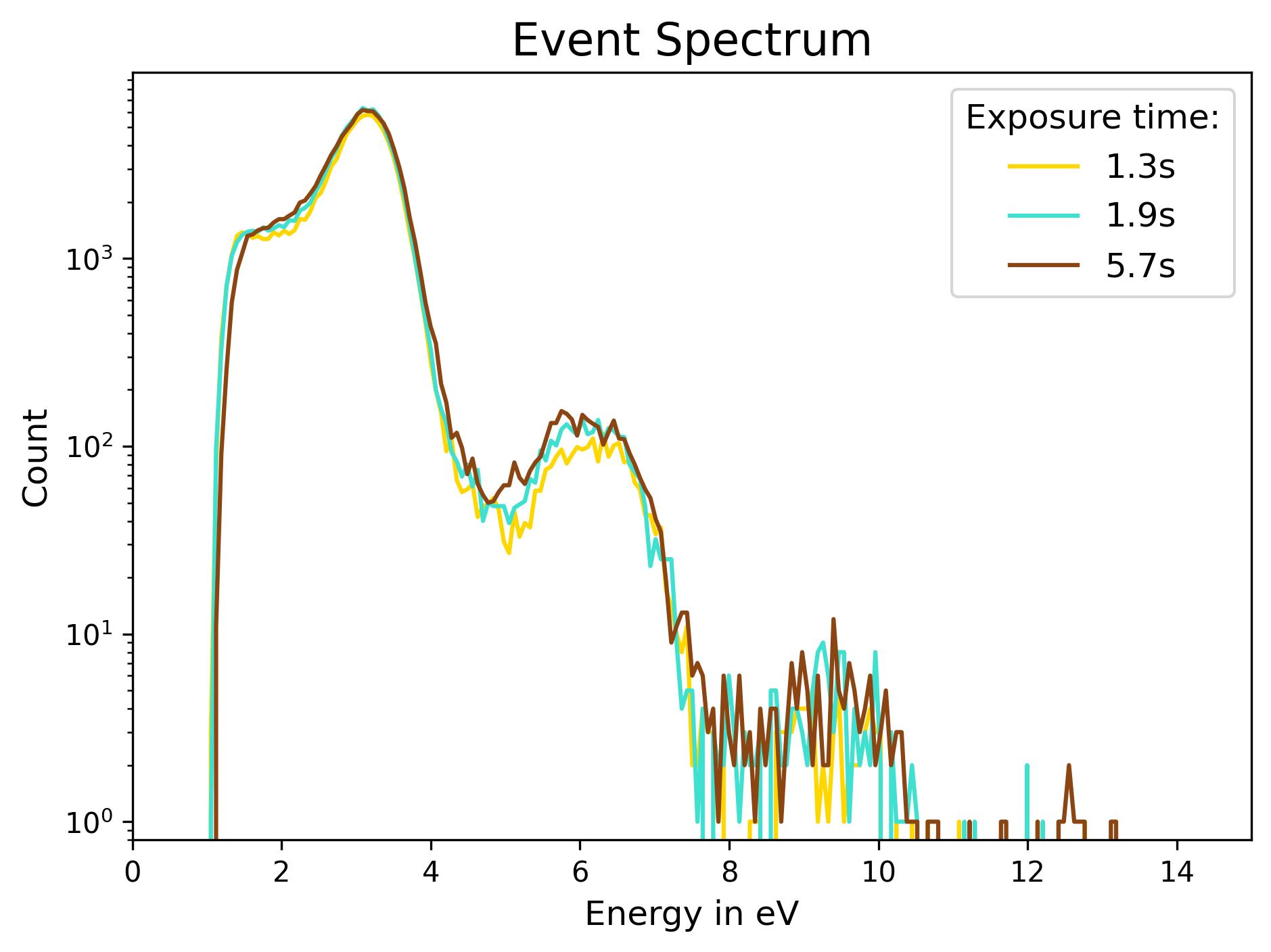}
    \caption{The non calibrated but filtered and clustered event spectra for the exposure sweep. The threshold for event filtering was 5 sigma.}
    \label{fig:ExposureHist}
\end{figure}

\begin{figure}[htb]
    \centering
    \includegraphics[width=0.55\linewidth]{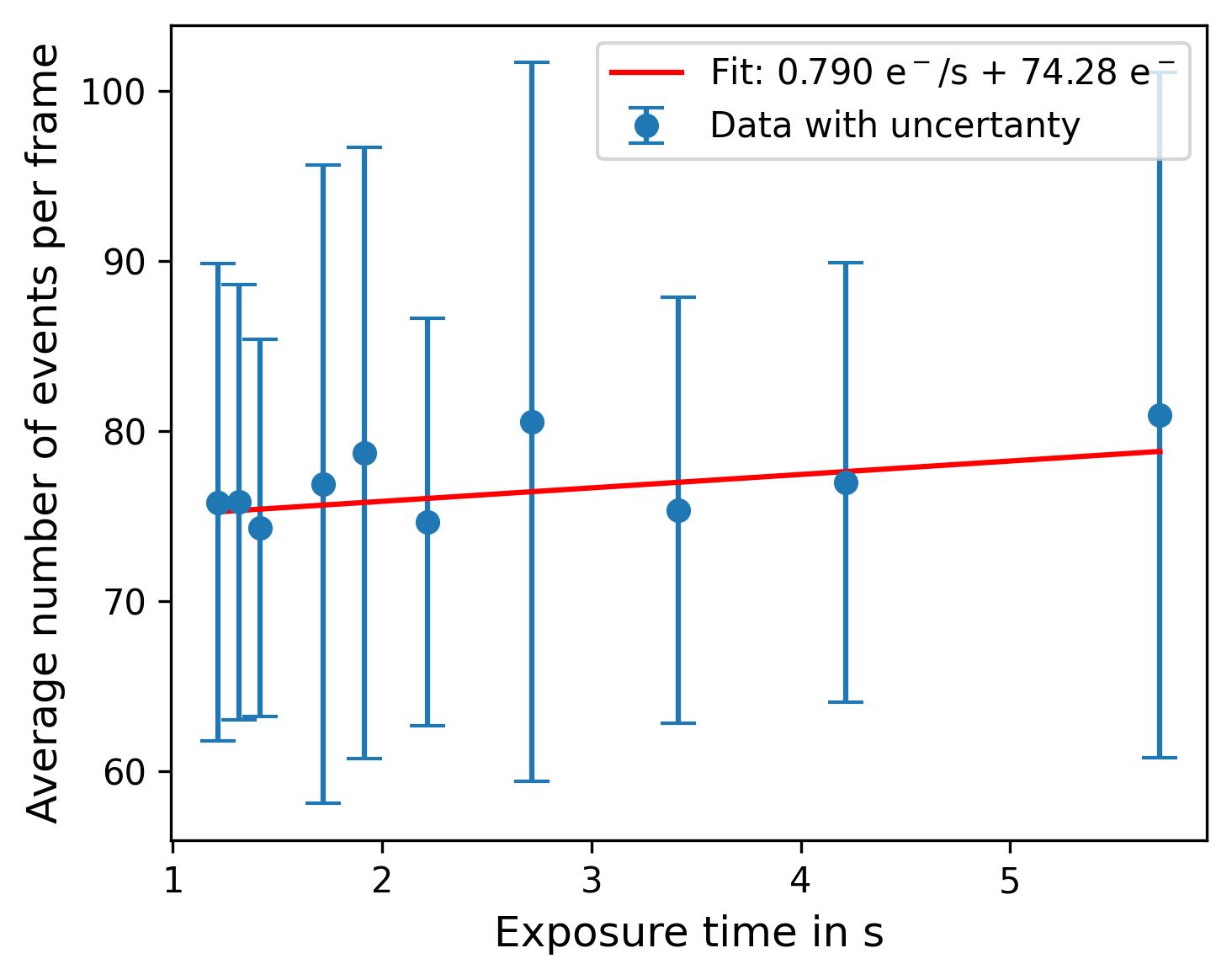}
    \caption{The average number of events detected over the whole sensor for the different exposure times. The red line shows the linear fit to the data points.}
    \label{fig:SEEMeanTime}
\end{figure}

The generation rate is determined by performing a linear fit to the pixel-averaged number of hits in the hit map. Multiple hits in one frame are treated as a single hit, since the fraction of multi-electron events is less than 2\% (\autoref{fig:SEEMeanTime}). The exposure-time-dependent generation or the slope of the linear fit --- commonly associated with the silicon bulk --- is distinguished from the exposure-time-independent or the offset component of the linear fit --- commonly associated with the sensor surface. In the bulk, thermally generated dark current is time dependent. At the surface, charges are generated by the readout process itself, mostly during charge transfer, and are therefore not exposure-time dependent. The spectra of filtered events before gain calibration for the different exposure times reveal a clear peak separation \autoref{fig:ExposureHist}.

For longer exposure times, more events are generated. This is observed for one-, two- and more-electron events without bias. 

\begin{figure}[htb]
    \centering
    \includegraphics[width=0.75\linewidth]{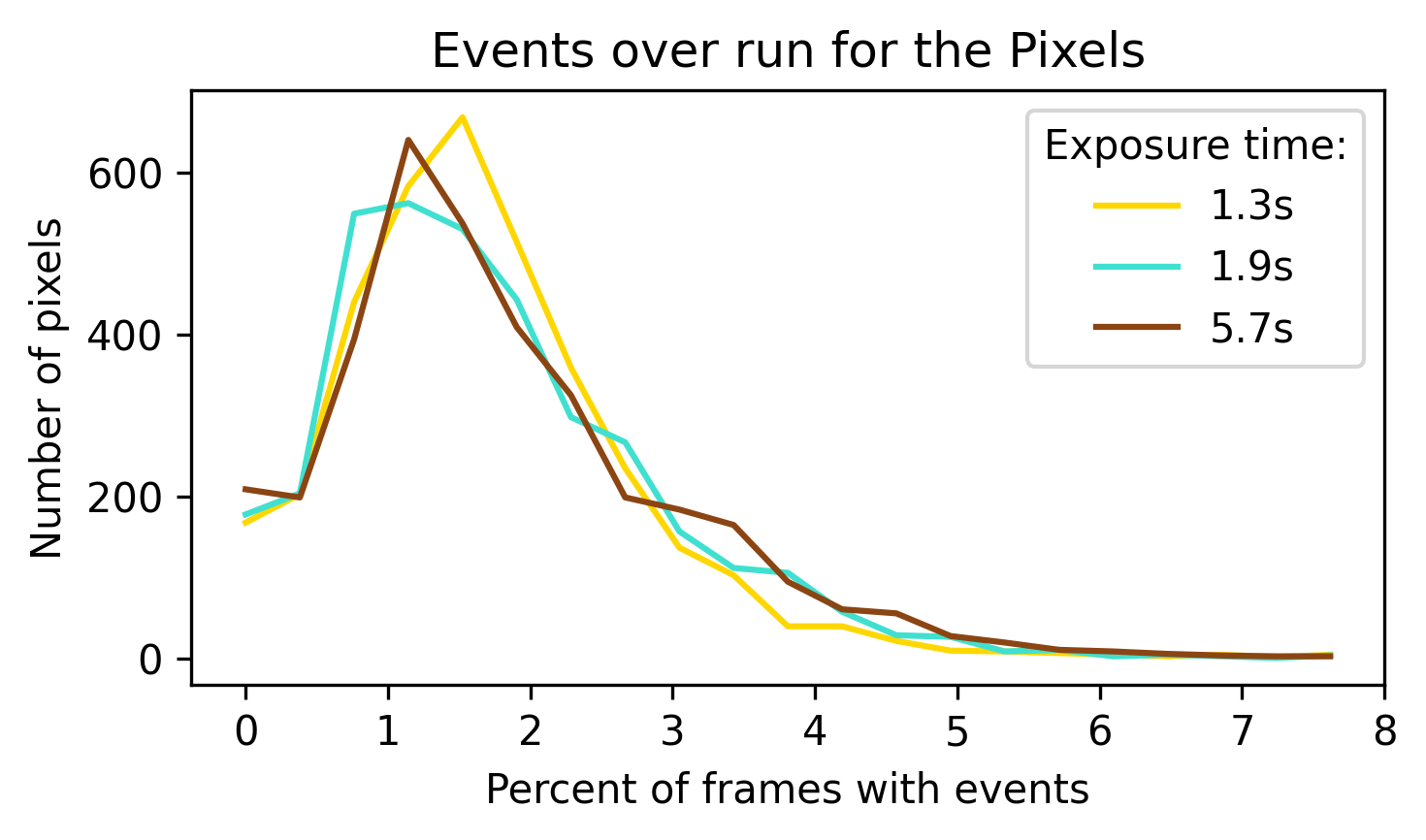}
    \caption{These histograms correspond to three different exposure times. For each exposure time, they show the number of pixels that recorded events in a given percentage of frames.}
    \label{fig:EventHist}
\end{figure}

The distribution of the number of events per pixel and for all frames, especially the two-dimensional histogram of event counts reveals a dominant contribution of about 75 events per readout that is independent of the exposure time \autoref{fig:EventHist}. This peak is higher for shorter exposure times and decreases for longer exposure times due to the additional bulk generation, which shifts pixels to higher event counts.

To determine the generation rate, the average number of pixels with a signal exceeding the 5-sigma threshold is calculated for each exposure run and plotted as a function of exposure time. For these data, clustered events are recombined, and events with two or three electrons are counted as a single hit. Since less than 2\% of pixels generate more than one charge, and fewer than 0.1\% generate more than two charges, this affects the results less than the quoted uncertainties. The data is fitted linearly, the results are shown in \autoref{fig:SEEMeanTime}.

The slope of the linear fit corresponds to the generation rate $R_{Gen}$, and the offset $R_\mathit{Off}$ is a time-independent component, with values of about:
\[ R_{Gen} = \SI{0,79 +- 0,44}{\electron \per \second } \]
\[ R_\mathit{Off} = \SI{74,3+-1,2}{\electron \per  frame} \]

After cuts, 3690 pixels are analysed. The pixel side length in DANAE is \SI{50}{\micro\meter}, and the thickness is \SI{450}{\micro\meter}. Converting the generation rate yields:

\[  R_{Gen} = \SI{18}{\electron \per pixel \per day} ~~ = ~~ \SI{7,0}{\electron \per \micro \gram \per day} ~~ = ~~  \SI{702e3}{\electron \per \cubic \centi \meter \per \second}\]

Literature values are available for the \textit{Sub-Electron-Noise Skipper-CCD Experimental Instrument} (SENSEI) detector, which was operated in a surface run with a pixel size of \SI{15}{\micro\meter}, a thickness of \SI{200}{\micro \meter}, and $624\times362$ pixels, for a total mass of \SI{71}{\milli\gram}. The sensor readout takes \SI{73,6}{\minute}~\cite{SenseiFirstConstrainsCrisler}. By scaling with pixel volume, comparable values of \SI{7,0}{\electron \per \micro\gram \per day} (DANAE) and \SI{10,8}{\electron \per \micro\gram \per day} (SENSEI) are obtained. However, this comparison only accounts for the exposure-time-dependent contribution. The offset of the linear fit or exposure-time-independent generation corresponds to \SI{74}{\electron \per frame} or \SI{7,7}{\electron \per \micro\gram \per frame} for DANAE, which is larger than expected and needs to be reduced in future studies and developments.

\section{Conclusion and outlook}
An exposure sweep was recorded with the prototype RNDR-DEPFET detector of DANAE, resulting in a time-dependent generation rate of \SI{18}{\electron \per pixel \per day}, which is comparable to literature values from semiconductor dark matter experiments. However, the time-independent generation rate was determined to \SI{74}{\electron \per frame}, which is larger than expected and is currently under investigation. This will be studied by assembling and characterizing additional detectors from a recently fabricated DEPFET production run developed by the semiconductor laboratory of the Max-Planck society, as well as by implementing a dedicated operation sequence that removes electrons from the internal gate before repetitive readout begins. This will also enable the investigation of exposure times shorter than the frame. 

For statistical purposes, the amount of collected data will be increased by longer measurement runs and will be expanded by exposure sweeps at different temperatures. That way, the ideal temperature can be determined for the experiment, as well as the expected temperature dependence of the exposure-time-independent charge-carrier generation. A light-emitting diode will be used for calibration in the upcoming measurement run. 

\FloatBarrier
\bibliography{mybibfile}

@phdthesis{Lauf,
    author = {Thomas Lauf},
    title = {Analysis and Operation of {DePFET} {X}-ray Imaging Detectors},
    school = {Technische Universität München},
    year = {2011}
}

@article{depfetBaehr,
author={B{\"a}hr, A.
and Kluck, H.
and Ninkovic, J.
and Schieck, J.
and Treis, J.},
title={{DEPFET} detectors for direct detection of {MeV} Dark Matter particles},
journal={The European Physical Journal C},
year={2017},
month={Dec},
day={26},
volume={77},
number={12},
pages={905},
issn={1434-6052},
doi={10.1140/epjc/s10052-017-5474-5},
url={https://doi.org/10.1140/epjc/s10052-017-5474-5}
}

@article{SenseiFirstConstrainsCrisler,
   title={{SENSEI}: First Direct-Detection Constraints on Sub-{GeV} Dark Matter from a Surface Run},
   volume={121},
   ISSN={1079-7114},
   url={http://dx.doi.org/10.1103/PhysRevLett.121.061803},
   DOI={10.1103/physrevlett.121.061803},
   number={6},
   journal={Physical Review Letters},
   publisher={American Physical Society (APS)},
   author={Crisler, Michael and Essig, Rouven and Estrada, Juan and Fernandez, Guillermo and Tiffenberg, Javier and Haro, Miguel Sofo and Volansky, Tomer and Yu, Tien-Tien},
   year={2018},
   month=8
}

@Article{cosinusAngloher,
    author={Angloher, G. and Carniti, P. and Cassina, L. and Gironi, L. and Gotti, C. and G{\"u}tlein, A. and Hauff, D. and Maino, M. and Nagorny, S. S. and Pagnanini, L. and Pessina, G. and Petricca, F. and Pirro, S. and Pr{\"o}bst, F. and Reindl, F. and Sch{\"a}ffner, K. and Schieck, J. and Seidel, W.},
    title={The {COSINUS} project: perspectives of a {NaI} scintillating calorimeter for dark matter search},
    journal={The European Physical Journal C},
    year={2016},
    month={8},
    day={08},
    volume={76},
    number={8},
    pages={441},
    issn={1434-6052},
    doi={10.1140/epjc/s10052-016-4278-3},
    url={10.1140/epjc/s10052-016-4278-3}
}

@misc{EssigDirect,
      title={Direct Detection of sub-{GeV} Dark Matter with Semiconductor Targets}, 
      author={Rouven Essig and Marivi Fernandez-Serra and Jeremy Mardon and Adrian Soto and Tomer Volansky and Tien-Tien Yu},
      year={2016},
      eprint={1509.01598},
      archivePrefix={arXiv},
      primaryClass={hep-ph}
}

@ARTICLE{DMRotation,
       author = {{van Albada}, T.~S. and {Bahcall}, J.~N. and {Begeman}, K. and {Sancisi}, R.},
        title = {Distribution of dark matter in the spiral galaxy {NGC} 3198.},
      journal = {The Astrophysical Journal},
         year = 1985,
        month = 8,
       volume = {295},
        pages = {305-313},
          doi = {10.1086/163375},
       adsurl = {https://ui.adsabs.harvard.edu/abs/1985ApJ...295..305V}
}

@Article{baehrSize,
	title={{First measurement results from DANAE - Demonstrating DePFET RNDR on a prototype Matrix}},
	author={Alexander Bähr and Holger Kluck and Peter Lechner and Jelena Ninkovic and Jochen Schiek and Hexi Shi and Wolfgang Treberspurg and Johannes Treis},
	journal={SciPost Phys. Proc.},
	pages={066},
	year={2023},
	publisher={SciPost},
	doi={10.21468/SciPostPhysProc.12.066},
	url={https://scipost.org/10.21468/SciPostPhysProc.12.066},
}

\end{document}